\documentclass[%
 aip,
 amsmath,amssymb,
 reprint,%
]{revtex4-1}

\usepackage{graphicx}
\usepackage{dcolumn}
\usepackage{bm}

\usepackage[utf8]{inputenc}
\usepackage[T1]{fontenc}
\usepackage{mathptmx}
\usepackage{etoolbox}
\usepackage{amsmath}

\usepackage{makecell}

\usepackage{hyperref}
\def\t#1{{\mathrm{#1}}}

\makeatletter
\def\@email#1#2{%
 \endgroup
 \patchcmd{\titleblock@produce}
  {\frontmatter@RRAPformat}
  {\frontmatter@RRAPformat{\produce@RRAP{*#1\href{mailto:#2}{#2}}}\frontmatter@RRAPformat}
  {}{}
}%
\makeatother
\begin{document}

\preprint{AIP/123-QED}

\title[Sample title]{Perspective on Nanoscaled Magnonic Networks}
\author{Qi Wang}
\affiliation{School of Physics, Huazhong University of Science and Technology, Wuhan, China}

\author{Gyorgy Csaba}%
\affiliation{P\'{a}zm\'{a}ny P\'{e}ter Catholic University, Faculty of Information Technology and Bionics, Budapest, Hungary
}%
\author{Roman Verba}
\affiliation{Institute of Magnetism, Kiev, Ukraine}

\author{Andrii V. Chumak}
\affiliation{Faculty of Physics, University of Vienna, Vienna, Austria}

\author{Philipp Pirro}
\affiliation{Fachbereich Physik and Landesforschungszentrum OPTIMAS, Rheinland-Pf\"{a}lzische Technische Universit\"{a}t Kaiserslautern-Landau, Kaiserslautern, Germany}

\date{\today}

\begin{abstract}

With the rapid development of artificial intelligence in recent years, mankind is facing an unprecedented demand for data processing. Today, almost all data processing is performed using electrons in conventional complementary metal–oxide–semiconductor (CMOS) circuits. Over the past few decades, scientists have been searching for faster and more efficient ways to process data. Now, magnons, the quanta of spin waves, show the potential for higher efficiency and lower energy consumption in solving some specific problems. While magnonics remains predominantly in the realm of academia, significant efforts are being made to explore the scientific and technological challenges of the field. Numerous proof-of-concept prototypes have already been successfully developed and tested in laboratories. In this article, we review the developed magnonic devices and discuss the current challenges in realizing magnonic circuits based on these building blocks. We look at the application of spin waves in neuromorphic networks, stochastic and reservoir computing and discuss the advantages over conventional electronics in these areas. We then introduce a new powerful tool, inverse design magnonics, which has the potential to revolutionize the field by enabling the precise design and optimization of magnonic devices in a short time. Finally, we provide a theoretical prediction of energy consumption and propose benchmarks for universal magnonic circuits.

\end{abstract}

\maketitle

\section{\label{sec:level1} Introduction}

Magnonics, a frontier field exploring spin waves in magnetically ordered materials, is advancing rapidly in  experimental and theoretical physics as well as in practical technology, leading the expansion of our understanding of magnetic interactions and their applications \cite{Mahmoud-2020-JAP, barman2021roadmap, pirro2021a, chumak2022advances}. This dynamic field is focused on the exploration and manipulation of information of various kinds: RF, analog, binary data and quantum entangled states, each requiring different processing methods. Spin waves offer many advantages that are crucial for technological progress. They enable profound miniaturization, allowing reductions to the nanometer scale in both wavelength and lateral dimensions of 2D and 3D elements, and operate over a wide frequency range – from one GHz to several THz. This versatility is enhanced by low-energy data transport and processing, the ability to precisely control properties through various physical parameters and phenomena, nanosecond-fast reconfigurability, diverse nonlinear and non-reciprocal phenomena, and the ability to use AI-powered inverse-design approaches. In addition, magnonic computing units are compatible with modern spintronic devices and can be integrated into CMOS electronics, e.g. using the chiplet approach. These advantages place spin waves at the forefront of innovations in information transport and processing.  

\begin{figure*}
\includegraphics[width=1\textwidth]{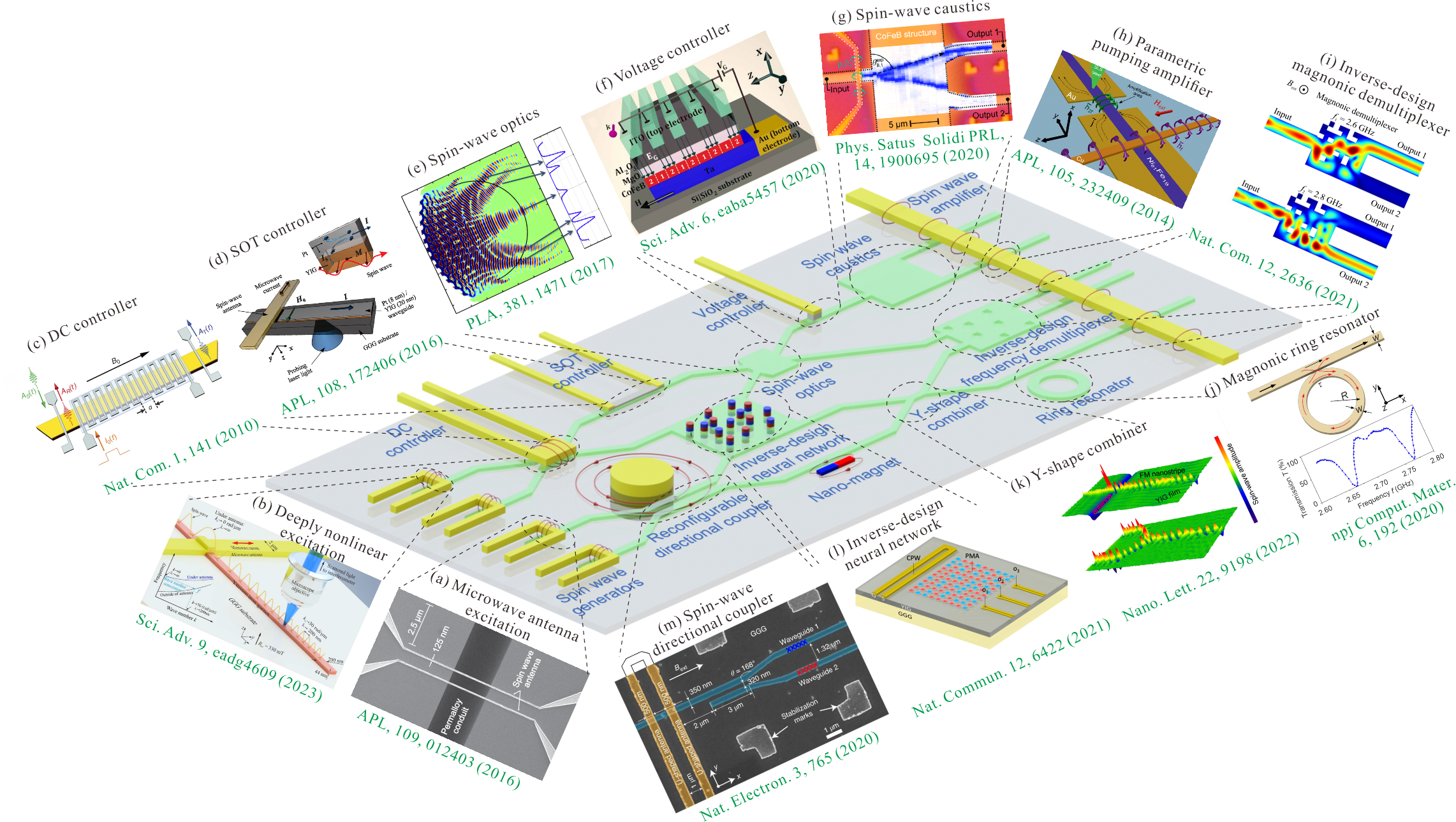}
\caption{\label{fig:1} A schematic of a nanoscale integrated magnonic network, including spin-wave generators, connectors, modulators and amplifiers with several key building blocks: (a) Microwave antenna excitation \cite{Ciubotaru-2016-APL}, (b) Deeply nonlinear excitation \cite{Wang-2023-SA}, (c) DC controller \cite{Chumak-2010-NC}, (d) SOT controller \cite{Evelt-2016-APL}, (e) Spin-wave optics \cite{Csaba-2017-PLA} (f) Voltage controller \cite{Choudhury-2020-SA}, (g) Spin wave caustics \cite{Heussner-2020-PRLRRL}, (h) Parametric pumping amplifier \cite{Bracher-APL-2014}, (i) Inverse-design magnonic demultiplexer \cite{Wang-2021-NC}, (j) Magnonic ring resonator \cite{Wang-2020-npj}, (k) Y-shape combiner \cite{Qin-2022-NL}, (l) Inverse-design neural network \cite{Papp-2021-NC}, (m) Spin wave directional coupler \cite{Wang-2020-NE}.}
\end{figure*}

The research field has significantly matured, resulting in a number of versatile prototype units suitable for cutting-edge data processing. At the same time, advances in the excitation, manipulation, amplification and guidance of spin waves have reached a level that allows the development and use of complex networks and circuits. An example of such a sophisticated magnonic network, with various essential building blocks, including spin-wave generators, connectors, modulators, and amplifiers, is conceptually depicted in Figure~\ref{fig:1} and is the central theme of this perspective article. Our primary focus is to explore the possibilities, address the challenges, and identify the yet-to-be-discovered components that are essential for moving magnonics from laboratory research to industrial application.

\section{Basic magnonic devices for data processing}

Binary data is a type of data that represents information using only two distinct values or symbols, typically denoted as "0" and "1", and is the cornerstone of modern computing and information technology. In complementary metal-oxide-semiconductor (CMOS) circuits, '0' and '1' respectively represent voltage levels. Compared to electron-based computing, wave-based computing provides an additional degree of freedom:information can be encoded in either amplitude or phase. Interference or diffraction effects between different waves provide a great opportunity to harvest the full potential of the wave nature for data processing and to simplify the design of wave-based computing devices. Spin waves are particularly well suited for wave-based computations due to their short wavelengths, high frequencies, low losses without Joule heating, and numerous nonlinear phenomena - nonlienearities are a must for any non-trivial computing. For this reason, spin-wave-based computing has attracted much attention from researchers during the last decade. 

\subsection{Spin-wave generation} 

The most common way to generate spin waves is to utilize the alternating Oersted field of a microwave antenna to locally induce magnetization precession as shown in Fig.~\ref{fig:1}(a) \cite{Ciubotaru-2016-APL}. For linear excitation, the width of the antenna limits the maximum wavenumber of the excited spin waves by the formula $k_{max}=2\pi/w$, where $w$ is the width of the antenna. Therefore, nanoscale antennas are needed to excite short spin waves. However, the spin-wave excitation efficiency of nanoantennas is greatly reduced due to the scaling and the increase in Ohmic resistance. To overcome this limitation, several other methods have been developed for the excitation of short-wavelength spin waves, such as magnonic grating couplers \cite{Liu-2018-NC}, magnetic vortex cores \cite{Sluka-2019-NN}, parametric pumping \cite{Sandweg-2011-PRL,Bracher-2017-PhysRep}, and geometry-induced wavenumber converters \cite{Stigloher-2018-APE}. Some of these methods have only been validated in unetched thin films, and the challenge of integrating them into nanoscale magnonic waveguides remains. Recently, a concise approach to excite short-wavelength spin waves by deeply nonlinear frequency shift has been realized, which can be easily integrated into magnonic circuits as shown in Fig.~\ref{fig:1}(b) \cite{Wang-2023-SA}. Although the excitation efficiency has improved over the past decades, it is still insufficient for large-scale industrial applications, especially at the nanoscale. One possible solution is to use magnetoelectric transducers, e.g. ones consisting of piezoelectric and magnetostrictive multilayers, which are driven by an AC voltage instead of an electric current and are therefore expected to be more energy efficient \cite{Cherepov-2014-APL}. 

\subsection{Spin-wave guiding and interconnection} 

Following the discussion of spin-wave excitation, it is logical to delve into the topics of spin-wave guiding and interconnection. The fundamental element for guiding spin waves is the magnonic waveguide, and the propagation properties of spin waves in the waveguides with widths ranging from millimeters to nanometers have been investigated \cite{Wang-2019-PRL,Demidov-2008-APL}. A multimode character has been observed in millimeter- or micrometer-wide waveguides raising the problem of scattering between different width modes if the spin waves pass through bends of the waveguides or impurities \cite{Clausen-2011-APL}. The easiest way to avoid this scattering is to reduce the width of the waveguide to the nanoscale, where the higher width modes are well separated in frequency due to the strongly quantized exchange interaction \cite{Heinz-2020-NL, Mohseni-2021-PRL}. Another advantage of nanoscale waveguides is their strong shape anisotropy, which makes the static magnetization parallel to the longitudinal axis of the waveguide even in the absence of an external magnetic field and makes it possible to realize zero-field spin wave propagation \cite{Wang-2018-SA,Nikolaev-2023-NL}. These self-biased magnonic waveguides are more suitable for complex 2D or 3D circuits where it is difficult to align the magnetization along the curved waveguide by simply applying a uniform external field. Therefore, nanoscaled magnonic waveguides are ideal spin-wave conduits for constructing 2D or 3D magnonic networks. 

Another challenge is to effectively interconnect different spin-wave channels to build a 2D or 3D magnonic network. Unfortunately, a simple X-type crossover has a significant drawback because it acts as a spin-wave re-emitter in all four connected spin-wave channels. The three-dimensional bridge connector used in modern electronic circuits is complex and expensive. Recently, a CoFeB/YIG bilayer Y-shape combiner has partially solved the reflection in the combiner region due to the nonreciprocal dispersion curve in the bilayer system as shown in Fig.~\ref{fig:1}(k) \cite{Qin-2022-NL}. However, information exchange between different waveguides cannot be realized using the nonreciprocal Y-shape combiner. Therefore, a passive connector consisting of a magnonic directional coupler based on the dipolar coupling between adjacent waveguides has been developed, as shown in Fig.~\ref{fig:1}(m) \cite{Wang-2020-NE}. One of the advantages of magnonic directional couplers is their excellent scalability -- the size can be reduced to a few tens of nanometers, which is much smaller than their microwave and photonic counterparts. Therefore, they can be used in magnonic logic circuits instead of 3D bridges to connect separate logic units. In addition, the magnonic directional coupler can also function as an analog microwave signal processing device, performing the functions of a tunable power splitter, frequency separator, or multiplexer, realizing the modulation of the spin-wave signal \cite{Wang-2018-SA,Wang-2020-NE}.

\subsection{Spin-wave modulators} 

At the heart of magnonic networks is data processing, which primarily involves magnon modulation. Spin waves can be effectively modulated by adjusting the magnetic material properties, including the applied external magnetic field, electric field, and spin current, or by the spin waves themselves through wave interference, diffraction, and nonlinearity. The most common method of spin-wave modulation is the electric current-controlled method, as shown in Fig.~\ref{fig:1}(c), in which the current-induced Oersted field locally shifts the dispersion curve and realizes the modulation of the spin-wave wavelength or group velocity \cite{Chumak-2010-NC}. Besides the current-induced Oersted field, the intrinsic property of the electron, the spin, can also be used to modulate magnons. Figure~\ref{fig:1}(d) shows the schematic of the Spin orbit torque (SOT) controller, in which a DC current flowing through the Pt layer (top) is converted into transverse spin accumulation by the Spin Hall effect  resulting in the modulation of the magnon propagation \cite{Evelt-2016-APL}. However, these methods are unsuitable for future low-power devices due to the high-power consumption caused by Joule heating, especially at the nanoscale. Similar to electric-field controlled spin-wave excitation, voltage can also be utilized to locally modulate magnetic anisotropy and thus manipulate spin-wave properties for energy-efficient nanoscale magnonic circuits, as shown in Fig.~\ref{fig:1}(f) \cite{Choudhury-2020-SA}.

As mentioned above, another advantage of spin-wave-based computation are the wave properties (interference and diffraction), nonlinearity, and anisotropic dispersion.  Figure~\ref{fig:1}(j) shows a magnonic resonator working as a notch filter, where the physical origin is also the wave interference \cite{Wang-2020-npj}. At the resonance frequencies, the output signal vanishes due to the destructive interference in the outgoing waveguide between the transmitted spin wave and coupled-back spin wave from the ring, and the resonance frequency depends on the input spin-wave power due to the nonlinearity. Finally, the anisotropy of the magnetic system was used to design a passive magnonic frequency demultiplexer \cite{Heussner-2020-PRLRRL}, as shown in Fig.~\ref{fig:1}(g), where caustic spin-wave beams have different beam angles corresponding to different spin-wave frequencies.

\subsection{Spin-wave amplifier}

So far, spin-wave generators, connectors and modulators have been implemented and discussed. However, an integrated magnonic circuit containing at least two individual magnonic units and which is also suitable for further cascading has not been realized experimentally. The main reason is that the spin-wave amplitude decreases after passing through the upper-level device due to damping preventing the next-level device to reach the required amplitude. The key element to solve this problem is a magnonic amplifier, which compensates for the loss and brings the spin-wave amplitude back to the initial state. The amplification of spin-wave signals can be realized by different mechanisms. In principle, spin-wave transistors could be used to generate large spin-wave signals \cite{Chumak-2014-NC}, similar to electronic amplifiers, but until now, no experiments on nanoscale magnonic transistors have been reported. Similar to the SOT controller, researchers also attempted to amplify the spin-wave signal using spin current. However, the direction of the external field and the composition of the materials must be precisely tuned to avoid the occurrence of nonlinear magnon scattering and auto-oscillations \cite{Evelt-2016-APL}. Recently, the research of Merbouche et al. has shed new light on this direction, achieving a true amplification effect where the output signal is larger than the input signal \cite{Merbouche-2023-arXiv}. Alternatively, spin wave amplifiers based on parametric pumping have been reported experimentally \cite{Bracher-APL-2014,Bracher-2017-PhysRep}, as shown in Fig.~\ref{fig:1}(h). In contrast to the direct excitation of spin waves, in so-called parallel parametric pumping, the component of the Oersted field is parallel to the direction of the static magnetization with twice the frequency of the spin wave. Once the propagating spin waves reach the pumping region, the microwave photons split into pairs of magnons at half the photon frequency to enhance the spin-wave signals. Very recently, Breitbach et al. \cite{Breitbach-2023-PRL} proposed a spin-wave amplifier based on rapid cooling, which is purely a thermal effect. The magnon system is brought into a state of local disequilibrium in excess of magnons. A propagating spin-wave packet reaching this region stimulates the subsequent redistribution process, and is, in turn, amplified. Although various magnonic amplifiers have been proposed, their further development is still an important research field in magnonics, especially since voltage-based spin-wave amplifiers with low power consumption are proposed only in theory \cite{Verba-2019-PhyRevApp} and need to be realized in experiments. 

In summary, several separate magnonic devices including spin-wave generators, connectors, modulators, and amplifiers (not only shown in Fig.~\ref{fig:1}) have been experimentally realized. However, the integration of these individual devices into a magnonic circuit remains a challenge due to the differences between the various methods, such as the need for a characteristic material system and a specific magnetic field direction or amplitude. In addition, improving the efficiency of spin-wave excitation, modulation, and detection is also an urgent issue to be addressed for future low-energy devices. Therefore, the focus of current research in the field of spin-wave-based computing should be to further optimize the performance of each individual magnonic device and, more importantly, to experimentally demonstrate magnonic integrated circuits with multiple interconnected magnonic devices.

\section{Neuromorphic magnonic networks, stochastic and reservoir computing}

Digital circuits employ a one-fits-all architecture and solve every computing problem by mapping it to a sequence of arithmetic operations. On the contrary, neuromorphic architectures are highly specialized and efficient analog processing tools, but applicable only to a particular problem type. A significant strength of the magnonics platform is that it can realize a wide variety of neuromorphic computing tasks and it naturally allows the interconnection of these units into a complex processing pipeline. 

For neuromorphic approaches, magnonic systems offer several unique advantages. Reprogrammability using external electric and magnetic fields or the domain state of the magnetic configuration (such as the spin ice-based reservoirs \cite{gartside2022reconfigurable}) is one of them. This way, magnonic systems may be tuned \cite{papp2021characterization}, e.g.,  to act as better reservoirs. In addition, a direct physical connection to the long-term magnetic memory units needed for training is straightforward since spin waves can be manipulated by static or dynamic coupling, e.g., to the free layer of a magnetic random access memory cell. In terms of connectivity, spin waves show a superior performance compared to CMOS in 2D since they can use the coherent coupling phenomena, which are the basis of directional couplers \cite{Wang-2018-SA, Wang-2020-NE} and frequency multiplexing \cite{Heussner-2020-PRLRRL} approaches. Full 3D magnonic systems are still in their infancy, but due to their wave nature and comparatively short wavelengths, very high interconnect density can be expected once the necessary patterning methods for magnonic materials are developed \cite{gubbiotti-2019-book}.

\subsection{Traditional Artificial Neural Network structures}

 The most obvious approach is to create a magnonic artificial neural network (mANN) in real space which consists of magnetic neurons connected by magnetic synapses, with information encoded in the amplitude or phase of spin waves. Various nonlinear magnonic devices can be used as neurons for this purpose, such as  magnonic bistabilities \cite{Wang-2023-SA} (Fig.~\ref{fig:1}(b)) or parametric amplifiers (Fig.~\ref{fig:1}(h))\cite{Bracher-APL-2014,Bracher-2017-PhysRep}. The latter can be used together with magnonic resonators \cite{Wang-2020-npj} to build analog magnonic adders\cite{Bracher-2018-JAP}. Waveguides, 2D or 3D magnonic structures can be used as synapses, the latter allowing frequency-selective addressing of neurons using spin-wave optical effects such as diffraction and caustic beams \cite{Heussner-2020-PRLRRL} (Fig.~\ref{fig:1}(g)). In advanced concepts of mANNs, the synapses and the neurons do not need to be physically separated as the non-linear waves themselves play the role of both\cite{Papp-2021-NC}. Training of the network requires a change of the synaptic strengths, i.e. of the spin-wave propagation in between the neurons. This could be achieved using the connection to traditional magnetic memories mentioned above. However, a promising alternative is the use of magnetic domain walls \cite{Wagner-2016-NN,Liu-2019-NN}, on the one hand, to reconfigure the spin-wave flux, e.g., in directional couplers. On the other hand, spin waves themselves can move domain walls \cite{Fan-2023-NatNano} and thus be used directly to write the memory as well. Such effects would enable unsupervised learning or online learning in an all-magnon circuit. 
 
 Electrical neural networks are the most often designed by computationally heavy machine learning functions - and can be trained for  a wide variety of pattern recognition classification tasks.  Machine learning algorithms were demonstrated to train linear and nonlinear interference of spin waves for a given task (Fig.~\ref{fig:1}(l)) \cite{Papp-2021-NC}- this possibly opens the door to many spin-wave-based neural computers. It is also possible to use frequency and phase encoding to recreate the structure of neural networks using spin waves and Ref. \cite{rahman2015wave} gives an example.

\subsection{$k$-space computing and reservoirs}
The fact that magnons are conveniently described in wave vector ($k$) space opens another interesting route, namely to create magnonic networks in  $k$-space. That is, a spatially small magnet in the micrometer range has many degrees of freedom in wave vector space, and these $k$-state states are straightforwardly accessible by GHz probing. One may look at the different $k$-states as highly interconnected computing nodes which are coupled by the pronounced and intrinsic magnon-magnon interaction -- and these interactions come 'for free', without the need for dedicated wiring. However, it is challenging to change these couplings selectively, so standard training schemes are difficult to apply to $k$-space computing. 

A workaround is to use reservoir computing (RC) approaches. RC uses an untrained physical system together with a simple trained linear classifier to do neural computation \cite{Nakajima-2020-JJAP}, while the complexity of the $k$-space dynamics naturally lends itself to  the implementation of the untrained layer. RC approaches have been successfully demonstrated using nanoscale magnetic components \cite{Koerber-2023-NatCom, Nakane-2018-IEEE_access} but also on larger systems  \cite{Watt-2021-PhysRevApp}. 

The scaling of the $k$-space approach in one single element is obviously limited. Thus, for networks with many computational nodes, a hybrid $k$-space/real-space architecture is likely needed, where $k$-space nodes (standing wave modes) and propagating magnons are simultaneously used. 

\subsection{Optically inspired neuromorphics}

Coherent optical processors can perform complex signal processing using (mostly linear) wave-interference \cite{ambs2010optical}. There are plenty of analogies between magnons and electromagnetic waves - so several ideas from the optical computing literature may be reused in the magnonic domains. The magnonic versions of some coherent optical processors have been demonstrated \cite{kiechle2023spin, Vogel-2015-NP} - these perform relatively simple signal processing functions (such as Fourier transform and Fourier-domain filtering, Fig.~\ref{fig:1}(e) \cite{Csaba-2017-PLA}) and are very useful at the input layer of neuromorphic processors.

\subsection{ONNs and RF neural networks}

Oscillatory Neural Networks (ONNs) use the phase and amplitude of interacting oscillators to perform analog computation. In magnonic systems various magnetic oscillators (STOs, SHNOs \cite{locatelli2014spin}) can be implemented that may act as neurons and magnonic couplings naturally act as synapses.

Spin waves may couple oscillatory computing elements together showing the possibility to realize oscillatory computing concepts. The oscillatory elements can be nanomagnets \cite{Zeng-2016-IEEE} or spin-torque oscillators \cite{macia2011spin}. A major bottleneck in artificial neural networks the limited 2D interconnectivity - which possibly could be alleviated by all to all magnonic interconnections \cite{grollier2020neuromorphic} \cite{kaka2005mutual} \cite{albertsson2021ultrafast}.

The inherently fast and low-power nature of spin dynamics could be especially useful for processing of radio-frequency signals. Spintronic neural networks to this end use electrical connections \cite{ross2023multilayer} - but magnonic interconnections could have an impact in the future.


\section{Magnonics for Computationally hard problems}

Computationally hard problems (such as NP-hard problems) are most often discussed in the context of quantum computing - but a large fraction of such problems do not require entanglement and can be handled either by classical interference or by probabilistic methods.

The work of Ref.\cite{Balynskiy-2020-arxiv} shows the application of spin wave interference (a.k.a holographic computing) for such quantum-inspired computing problems. The  work of Ref. \cite{Khivintsev-2016-JAP} gives a proof of principle demonstration of prime number factorization.

Using  the analogy between photonic and magnonic systems led to the demonstration of a magnonic version of coherent Ising machines \cite{litvinenko2023spinwave}. This device uses magnons for precise, short-term storage of analog information while the magnon state is manipulated by electrical circuitry. 

Random thermal fluctuations  always play  an important role in the behavior of room-temperature magnetic systems - and probabilistic computing algorithms may take advantage this abundance of free random number generators. One example are parametrons for probabilistic/stochastic computing: \cite{Makiuchi-2021-APL}, which generate a stable magnetic oscillation with random phase. In spintronics, the low-energy barrier MTJs are used as $p$-bits for building probabilistic computing hardware. Exploiting randomness is intensely researched in spintronic devices, but yet yet explored in pure magnonic systems.










\section{Machine learning-based inverse design as a new tool for magnonics}

As described in the previous sections, many magnonic devices have been demonstrated. However, the development of each of these devices requires specialized effort and complex investigations. The concepts of machine learning-based inverse design are well established in photonics and for the automated design of large-scale integrated CMOS circuits. Their application to magnonics is a more recent development. This incorporation brings groundbreaking opportunities to the field of magnonics, where the influence of machine learning and inverse design is expected to grow exponentially in the coming times.

Many magnonic computing units are discussed in this perspective article and shown in Fig.~\ref{fig:1}, each typically suited for a single function. The inverse-design method allows functionalities to be specified first, and a feedback-based (e.g., machine learning) computational algorithm is used to design a device with the desired functionalities. Two reports on the successful use of inverse design in magnonics have been published in parallel. To demonstrate the universality of this approach, linear, nonlinear, and nonreciprocal magnonic functionalities were studied in Ref. \cite{Wang-2021-NC}, and the same algorithm was used to create a magnonic (de)multiplexer, a nonlinear switch and a Y-circulator. A three-port proof-of-concept prototype is based on a rectangular ferromagnetic domain that can be patterned using square voids, and a direct binary search algorithm was used – see Fig.~\ref{fig:1}(i). The authors in Ref. \cite{Papp-2021-NC} took a more complex functionality and (inverse) designed a neural network in the form of a YIG domain with an array of nanomagnets deposited on top (see Fig.~\ref{fig:1}(l)). It is shown that all neuromorphic computing functions, including signal routing of a nonlinear activation, can be performed by spin-wave propagation and interference. In particular, vowel recognition is used to demonstrate the inverse-design magnonic neural network.

Recently, a spin-wave lens was designed using machine learning and experimentally demonstrated \cite{Kiechle-2022-lens}. A landscape in the saturation magnetization of a YIG film, realized by direct focused-ion-beam (FIB) irradiation, was used to train the network using a custom micromagnetic solver SpinTorch (https://github.com/a-papp/SpinTorch) with built-in automatic gradient calculation and performing backpropagation through time for spin-wave propagation. The focusing of the spin-wave beam by an inverse-designed lens was successfully demonstrated \cite{Kiechle-2022-lens}, proving the feasibility of the approach for any other complex functionality.

A different inverse design approach was used to generate linear spin-wave pulses with a desired spatial-temporal profile in magnonic waveguides recently \cite{casulleras2023pulses}. The problem is that the development of fast magnonic information-processing devices requires operating with short spin-wave pulses, but, the shorter the pulses, the more affected they are by information loss due to broadening and dispersion. The capability of inverse engineering spin-wave pulses and controlling their propagation allowed to solve this problem \cite{casulleras2023pulses}. 

The conceptual advantages of inverse design include: (i) The universality of inverse design - it is a tool that can be successfully used for RF applications, for binary and neuromorphic computing, as well as for the design/optimization of any component of any magnonic network. (ii) Inverse design has the potential to allow combining different functionalities in the same device - e.g., an RF multiplexer can be combined with a Y-circulator and a power limiter.  (iii) The complexity of the functionality is limited only by the size of the parameter matrix and by the number of states in a single matrix element - e.g. instead of cascading two half adders to realize the full adder, the same chip (potentially of the same size) but with the denser matrix can be used. (iv) Inverse design is naturally applicable to the realization of neuromorphic circuits allowing to perform very complex tasks.

The main challenges include: (i) The large computational and memory requirements that limit the complexity of the inverse-design devices. The solution under development is to solve an adjoint system of LLG equations during the gradient computation, thus reducing considerably the memory consumption.  (ii) Any optimization problem is constrained by the system of PDEs, and the computation of gradients remains a significant hurdle in optimization, leading to a high degree of freedom after discretization  \cite{chumak2022advances}. However, the gradient of such a problem can be efficiently computed by solving the adjoint problem to the constraining PDE \cite{bruckner2017solving}. Furthermore, the extension of this approach to micromagnetics operating in frequency domain \cite{perna2022computational} will significantly expand the novel field of inverse magnonics. (iii) Typically, a magnonic device is first designed numerically and then fabricated \cite{Kiechle-2022-lens}. The development of experimental systems that allow for large reconfigurable matrices with a significant number of elements and a large number of states in each of them would allow for fully experimental inverse-design magnonics.

\section{Benchmarks of universal magnonic circuits }

Energy consumption is one of the most crucial characteristics of future information processing devices. Magnonics offers great perspectives in the development of devices with ultralow power consumption, provided by low inherent magnetic losses in magnetic materials, especially in dielectric ones. 
Here we consider a general approach for the estimation of consumption, as well as scaling rules of other important characteristics. 

Energy of a SW pulse of the duration $\tau$ and amplitude $c_k$ (defined as a canonical amplitude) is equal to $E=M_s S v_\t{gr}\tau \omega_k |c_k|^2/\gamma$, being dependent on the saturation magnetization $M_s$, waveguide (SW bus) cross-section $S$, pulse velocity $v_\t{gr}$ and carrier frequency $\omega_k$, and gyromagnetic ratio $\gamma$. In an amplitude-coded logic a part of the energy is lost during some of the processing operations, e.g. in a half-adder operation ``1''+``1'' $\to$ ``0''(sum) + ``1''(carry). Thus, consumption can be reasonably estimated as the energy of a single SW pulse. 

Most of the developed all-magnon logic gates rely on 4-magnon nonlinearity, characterized by nonlinear frequency shift $T_k$, $\Delta\omega_\t{NL} = T_k |c_k|^2$. Correct processing requires the pulse spectrum width ($\sim\tau^{-1}$) to be much less than $\Delta\omega_\t{NL}$, i.e. $\tau \sim C/|T_k c_k^2|$ with the coefficient being of the order of $C \sim 10$ (accurate condition depends on a logic gate design \cite{Wang-2020-NE}). Thus, we get the minimal consumption estimation $E = C M_s S v_\t{gr}\omega_k/(\gamma|T_k|)$. It is independent of the SW amplitude, since a larger SW amplitude allows for the reduction of SW pulse duration. 

In the limit of short exchange-dominated SWs $\omega_k\approx -T_k \approx \omega_m \lambda^2 k^2$ ($\lambda$ is the exchange length), so $E \approx 2C M_s S \lambda \sqrt{\omega_M \omega_k}/\gamma$. The consumption increases with SW carrier frequency -- it is an inevitable general rule for any physical mechanisms -- but quite slowly. We see that the most feasible way to reduce consumption is the fabrication of smaller SW buses and playing with material parameters; more involved approaches are discussed below. 

Maximal SW frequency and, thus, minimal SW wavelength $\lambda_\t{SW}$, are other crucial characteristics. As discussed above, they are mostly determined from the single-mode operation condition \cite{Wang-2018-SA}, and depend on SW bus width, $\lambda_\t{SW,min} \approx w/2$ in the exchange limit, although dipolar contribution can somewhat alter this rule. Still, the smaller the waveguide, the larger the carrying frequency which is accessible. The operation frequency is inversely proportional to the pulse duration, $f = (\tau + \tau_\t{r})^{-1}$, and, thus, can be enhanced by increasing of the SW amplitude. The pulse separation $\tau_\t{r}$ is a more involved characteristic, given by dispersion smearing of a pulse and relaxation of magnetization oscillations in a processing unit. The last one is important for resonator-based gates only (e.g., ring resonator \cite{Wang-2020-npj}), and could be quite large in materials with low Gilbert damping $\alpha_G$: $\tau_\t{r} \sim 1/(\alpha_G \omega_k)$.     

\begin{table}
\begin{center}
\begin{tabular}{|c|c|c|c|c|c|}
 \hline
   Material 	       		  & YIG$^\t{a}$  & YIG & CoFeB & Ga:YIG & CMOS$^\t{b}$ \\ \hline
   \makecell{Cross-section, \\ $t\times w$ (nm)}& $30\times 100$ & $20\times 50$ & $5\times 20$ & $20\times 50$ & (7 nm)$^\t{c}$ \\ \hline
   \makecell{Max. SW frequency \\ (GHz)}  & 2.28 & 8.4 & 44 & 21 & -- \\ \hline
   Min. wavelength (nm)		  & 340 & 100 & 35 & 100 & -- \\ \hline
   Velocity, $v_{gr}$ (m/s)	  & 25 & 1150 & 1950 & 3800 & -- \\ \hline
   Consumption (aJ/bit) 	  & 24.6 & 10 & 14 & $\sim$0.5 & 35.3 \\ \hline
   \makecell{Max. operation \\ frequency$^\t{d}$ (MHz)} & 8 & 170 & 900 & 420 & $\sim 3000$\\ \hline
   \makecell{Relaxation \\ frequency (MHz)}& 2.5 & 10 & 1500 & 80 & -- \\ \hline
   Area ($\t{\mu m}^2$)		  & 5.58 & -- & -- & -- & 1.024 \\ \hline
\end{tabular}
\end{center}
\caption{Benchmarking of 4-magnon interaction based magnonic logic devices and its comparison to 7-nm CMOS technology; calculations are made for bias-free case. $^\t{a}$ Data for a particular half-adder design from \cite{Wang-2020-NE}. $^\t{b}$ The values are calculated using Cadence Genus for 7-nm CMOS technology \cite{Wang-2020-NE}. $^\t{c}$ CMOS transistor gate size. $^\t{d}$ Calculated exemplary for $|c_k|^2 = 0.1$.}\label{t:benchmark}
\end{table}

Estimation of the considered characteristics for popular magnonic materials are given in Table~\ref{t:benchmark} together with accurately defined characteristics of a particular magnonic half-adder design \cite{Wang-2020-NE}, and are compared to promising 7-nm CMOS technology. One can see that technologically reliable YIG($30\times100\,$nm)-based half-adder already yields lower consumption than CMOS, and SW bus decrease and utilization of other materials allows to reach even smaller values. A promising in terms of energy efficiency way could be utilization of low-$M_s$ ferrimagnetics, like Ga:YIG, which may allow for the consumption only one order higher than the thermal stability limit of a pulse ($E \sim (10-20)k_B T \sim 0.04-0.08\,$aJ). Additional benefit of Ga:YIG is bias-free perpendicular magnetization state, in which nonlinear SW stability is much enhanced \cite{Wang-2023-SA}.  But one should be aware of the problem of excitation and reading of SWs in low-$M_s$ materials, as well as that close to the compensation point $M_s \to 0$ the mentioned estimation is incorrect.

It should be noted, that the calculated consumption accounts for the magnonic system only. Ohmic losses in conventional conducting wire transducers can overcome intrinsic losses by several orders and are detrimental for practical use. The best solution to date is the utilization of magnetoelectric effects \cite{Fiebig_JPD2005, Nikonov_ProcIEEE2013} to excite and process SWs, e.g. voltage-controlled magnetic anisotropy effect \cite{Verba-Book2017, Rana-2019-CP} or piezoelectric-magnetostrictive cells, which turns total consumption to aJ level \cite{Wang-2020-NE}. While not all ferromagnetics demonstrate efficient magnetoelectric effects 
development of hybrid structures could be a viable solution \cite{Au_PRAppl2023}.

The processing frequency of magnonic logic devices is overall smaller compared to CMOS, especially if a gate design requires relaxation time. A game-changing here could be the utilization of antiferromagnetics (AFMs), where SW frequency and velocity are much enhanced compared to ferromagnets \cite{Baltz-2018-RMP}. However, operational principles for AFM magnonics should be developed from the very beginning, as many ones developed for ferromagnetic are not applicable to AFMs.

Simultaneously, it should be noted that the frequency of several GHz is the clock speed in CMOS circuits. Complex processing operations require many subsequent simple logic operations, while a variety of nonlinear phenomena allows one to realize the same operations in magnonics by much less units. This difference is expected to be especially pronounced in neuromorphic and other non-Boolean computing systems, making magnonics circuits to be potentially not just more energy-efficient, but also faster and simpler in the architecture.  

Future improvement can be imagined in different ways. This could be the development of phase-coded logic. Although many phase-coded gates have been developed \cite{Fischer-2017-APL}, they still have the same problem of unnormalized output, i.e. loss of a part of the signal. Utilization of other nonlinear mechanisms is also a promising pathway as well. This includes utilization of other 4-magnon nonlinearity than simple frequency shift and searching for anomalously enhanced 4-magnon nonlinearity. Lower-order, 3-magnon nonlinearity is also intriguing -- in this case $E_\t{min} \sim |c_k|$ and could be potentially reduced to thermal limits. Unconventional 3-magnon processing with localized SWs was realized in \cite{Korber-2023-NC}, while processing of propagating SWs is an uncovered direction. Finally, magnetic solitons, like propagating SW bullets, could be a promising way, especially for the reduction of the pulse (soliton) duration. To date, this topic is almost uncovered (please, distinguish dynamic magnetic solitons from SW envelope solitons, which are a kind of nonlinear SW pulse used in standard magnonics).

\section{Concluding remark}

The purpose of this perspective article is to list the main components of future fully functional complex magnonic networks, to discuss their main advantages and limitations in terms of integrability, and to share our vision of the main challenges and prospects in the field of magnonics in a concise form. To summarize, the number and scope of recent breakthroughs is quite impressive, as Fig.~\ref{fig:1} shows only a small selection of examples \cite{Mahmoud-2020-JAP, barman2021roadmap, pirro2021a, chumak2022advances}. The range of magnetic materials, nanoscale hybrid structures, physical phenomena involved, information processing concepts and types of information considered by the modern field of magnonics is also exceptional. This forms a solid technological, methodological and scientific basis for the further development of integrated magnonic networks, and for the field to deliver industrially competitive devices to occupy their niches in our modern life.

\begin{acknowledgments}
Q.W. acknowledges support from the National Natural Science Foundation of China, the startup grant of Huazhong University of Science and Technology (Grant no. 3034012104). R.V. acknowledges support by National Research Foundation of Ukraine, Grant \#2020/02.0261. A.C. acknowledges support by the European Research Council project ERC Proof of Concept  No. 271741898 "5G-Spin" and by the Austrian Science Fund (FWF) through Project No. I 4917-N "MagFunc". P.P. acknowledges support by the European Research Council Starting Grant No. 101042439 "CoSpiN".
\end{acknowledgments}

\section*{Data Availability Statement}

AIP Publishing believes that all datasets underlying the conclusions of the paper should be available to readers. Authors are encouraged to deposit their datasets in publicly available repositories or present them in the main manuscript. All research articles must include a data availability statement stating where the data can be found. In this section, authors should add the respective statement from the chart below based on the availability of data in their paper.

\bibliography{Bibliography}

\end{document}